# Comments to DØnote 4760-CONF

S. Sultansoy

Gazi University, Faculty of Arts and Sciences, Department of Physics, 06500 Teknikokullar, Ankara, TURKEY
and
Institute of Physics, Academy of Sciences, H. Cavid Avenue 33, Baku, AZERBAIJAN

**Abstract**

A brainstorming on recent DØ data on a search for the Higgs boson in H $\rightarrow$ WW$^{(*)}$ $\rightarrow$ l$^+$νl'$^-$ν (l, l' = e, μ) decays is performed. As a result, it is possible that the data favors $m_H$ = 160 GeV if there are four SM families.

It is known that the existence of the fourth SM family can give opportunity for Tevatron to observe the Higgs boson before the LHC (see [1] and references therein). In this note I discuss recent DØ data on a search for the Higgs boson in H $\rightarrow$ WW$^{(*)}$ $\rightarrow$ l$^+$νl'$^-$ν (l, l' = e, μ) decays [2].

Number of signal (3 SM families) and background events expected and number of events observed with the Run II DØ detector from April 2002 to June 2004 ($L^{int} \approx 300$ pb$^{-1}$) is presented in Table I, which is depicted from Table II of Ref [2]. Deviations of number of events observed from expected background are added in the last column of the table. Let me mention significant negative deviations at $m_H$ = 100, 120 and 200 GeV as well as essential positive deviation at $m_H$ = 160 GeV. Naturally, in the case of three SM families signal events do not effect situation.

Table I. Number of signal (3 SM families) and background events expected and number of events observed. Only statistical uncertainties are given.

| $m_H$, GeV | H $\rightarrow$ WW$^{(*)}$ | Background sum | Data | Deviation |
|---|---|---|---|---|
| 100 | 0.007 ± 0.001 | 31.0 ± 2.8 | 25 | -2.1 σ |
| 120 | 0.11 ± 0.01 | 30.2 ± 2.5 | 23 | -2.9 σ |
| 140 | 0.33 ± 0.01 | 21.4 ± 1.6 | 21 | -0.25 σ |
| 160 | 0.54 ± 0.02 | 17.7 ± 1.0 | 20 | +2.3 σ |
| 180 | 0.36 ± 0.01 | 19.1 ± 1.0 | 20 | +0.9 σ |
| 200 | 0.17 ± 0.01 | 20.0 ± 1.1 | 16 | -3.6 σ |

In principle, negative deviations can be caused from overestimation of the background, e. g. as a result of overestimated luminosity. The uncertainty on the luminosity measurement is 6.5 % [2]. Number of signal (4 SM families) and background events reduced by 6.5% are given in Table II. It is seen that numeric values of negative deviations at $m_H$ = 100, 120 and 200 GeV are reduced essentially, whereas positive deviation at $m_H$ = 160 GeV becomes more significant, namely, 3.9 σ. And this deviation coincides with signal expected at $m_H$ = 160 GeV if there are four SM families. Of course, this coincidence may be accidental. However, it is quite possible that DØ data give us a clue relating to Higgs boson.

Table II. Number of signal (4 SM families) and background events expected (reduced by 6.5%) and number of events observed.

| $m_H$, GeV | H → WW$^{(*)}$ | Background sum | Data | Deviation |
|---|---|---|---|---|
| 100 | 0.05 ± 0.01 | 29.0 ± 2.6 | 25 | -1.5 σ |
| 120 | 0.82 ± 0.08 | 28.2 ± 2.3 | 23 | -2.3 σ |
| 140 | 2.45 ± 0.08 | 20.0 ± 1.5 | 21 | +0.7 σ |
| 160 | 4.04 ± 0.15 | 16.5 ± 0.9 | 20 | +3.9 σ |
| 180 | 2.69 ± 0.08 | 17.9 ± 0.9 | 20 | +2.3 σ |
| 200 | 1.27 ± 0.08 | 18.7 ± 1.0 | 16 | -2.7 σ |

The correctness of our interpretation will be checked within a few months, because the collected luminosity per experiment (DØ and CDF) already has exceeded 1 fb$^{-1}$. This means 13.0±0.5 signal events per experiment comparing to 55±3 background events at DØ.

With $m_H$ ≈ 160 GeV, W bosons are produced practically at rest. Therefore, the energies of leptons from signal events are ≈ 40 GeV. In principle, this property can be used to reduce background.

Another promising decay chain is H → WW$^{(*)}$ → lvjj which has 7 times larger branching comparing to pure leptonic mode. At $m_H$ ≈ 160 GeV the signal events have clear signature: back-to-back jets with $E_{j1}$ ≈ $E_{j2}$ ≈ 40 GeV plus charged lepton with $E_l$ ≈ 40 GeV plus missing energy.

**References**

[1] E. Arik et al., hep-ph/0502050 (5 Feb 2005).
[2] DØ Collaboration, "Search for the Higgs boson in H → WW$^{(*)}$ → l$^+$vl'$^-$v (l, l' = e, μ) decays in p$\bar{\text{p}}$ collisions at $\sqrt{s}$ = 1.96 TeV", DØnote 4760-CONF (March 14, 2005).